\documentclass[conference]{IEEEtran}
\IEEEoverridecommandlockouts
\usepackage{cite}
\usepackage{amsmath,amssymb,amsfonts}
\usepackage[linesnumbered]{algorithm2e}
\usepackage{algpseudocode}
\usepackage{algcompatible}
\usepackage{graphicx}
\usepackage{textcomp}
\usepackage[dvipsnames]{xcolor}
\usepackage{textcomp}
\usepackage{hyperref}
\usepackage{subfiles} 
\usepackage{tabto}

\def\BibTeX{{\rm B\kern-.05em{\sc i\kern-.025em b}\kern-.08em
    T\kern-.1667em\lower.7ex\hbox{E}\kern-.125emX}}
    
\begin{document}
\bstctlcite{IEEEexample:BSTcontrol}

\title{Design Space Exploration of Approximate Computing Techniques with a Reinforcement Learning Approach
\\
\thanks{This work has received funding from the APROPOS project in the European Union’s Horizon 2020 research and innovation program under the Marie Skłodowska-Curie grant agreement No 956090 and from the project "National Center for HPC, Big Data and Quantum Computing", CN00000013 (Bando M42C – Investimento 1.4 – Avviso Centri Nazionali” – D.D. n. 3138 of 16.12.2021, funded with MUR Decree n. 1031 of 17.06.2022). \textbf{}}
\thanks{}
}

\author{\IEEEauthorblockN{Sepide Saeedi, Alessandro Savino, Stefano Di Carlo}
\IEEEauthorblockA{\textit{Politecnico di Torino, Control and Computer Eng. Dep., Torino, Italy}
 \\
emails: \{sepide.saeedi, alessandro.savino, stefano.dicarlo\}@polito.it}
}

\maketitle

\begin{abstract}
Approximate Computing (AxC) techniques have become increasingly popular in trading off accuracy for performance gains in various applications. Selecting the best AxC techniques for a given application is challenging. Among proposed approaches for exploring the design space, Machine Learning approaches such as Reinforcement Learning (RL) show promising results. In this paper, we proposed an RL-based multi-objective Design Space Exploration strategy to find the approximate versions of the application that balance accuracy degradation and power and computation time reduction. Our experimental results show a good trade-off between accuracy degradation and decreased power and computation time for some benchmarks.
\end{abstract}


\section{Introduction}

Approximate Computing (AxC) techniques have become increasingly popular to improve the energy efficiency of computations, power consumption, and computation time, degrading the accuracy of the computation. However, the reduction is possible by sacrificing the computation accuracy at different parts of the computing chain. Trading off accuracy for improved power consumption and performance is known as Approximate Computing (AxC)~\cite{Bosio:2022aa}. 
Selecting the most suitable AxC techniques for an application is challenging and many publications proposed how to explore the design space to find the most suitable AxC techniques for an application~\cite{Hu:2019aa}. Among various approaches, such complex Design Space Exploration (DSE), including genetic algorithms and simulated annealing~\cite{Rizakis:2018aa}, machine learning (ML) ones such as Reinforcement Learning (RL) showed promising results in improving the DSE~\cite{Wu:2021aa}, minimizing the number of designs to evaluate while maximizing the quality of the DSE model and reducing the exploration time~\cite{gautier:2022sherlock}.

RL is a machine learning paradigm in which an agent learns a desired behavior by interacting with a dynamic environment. A standard RL setup involves an agent connected to its environment via observations and actions. During each interaction with the environment, the agent receives observations about the current state of the environment and a reward based on its previous action. Based on this, it must select a new action to move toward an optimization goal, learning how to do this over time by systematic trial and error~\cite{kaelbling1996rl}.

Wu et al. proposed a DSE framework using RL to optimize resource allocation and critical path timing~\cite{Wu:2021aa}. The framework extracts the data flow graphs from the HLS C/C++, then the RL-based DSE engine explores the resource allocation options and finds optimized or Pareto solutions. The results show that their proposed RL-based engine outperforms genetic algorithms and simulated annealing. However, the exploration is limited to the trade-offs of different resources and critical path timing on an FPGA for a specific approximated version of an application, not considering the different approximated versions. A similar ML-based approach to DSE was proposed in~\cite{gautier:2022sherlock} to reduce the exploration time of the HLS tools for a certain application. 

In~\cite{Savino:2019aa}, the authors propose a new approach for the DSE of approximate applications to minimize the error rate using the data lifetime to select the approximation. Though their experimental results show their approach's effectiveness, the paper does not cover multi-objective optimization, i.e., simultaneously considering accuracy, power consumption, and computation time.

This paper proposes an RL-based multi-objective DSE methodology based on the resource selection process proposed by~\cite{Savino:2019aa} able to optimize an application balancing the final accuracy with the power consumption and the computation time introduced by the selective activation of approximate adders~\cite{Mrazek:2017aa} and multipliers~\cite{Ceska:2017aa}. 

\section{Methodology}
\label{sec:methodology}

This paper considers a CPU running software with dedicated instructions to trigger different approximate adders and multipliers. To generate approximate versions of a target application, the strategy is to select variables from the target application and approximate all sums or multiplications on those variables (as in~\cite{Savino:2019aa}). Since several design choices exist (i.e., different sets of variables and different approximate adders and multipliers), we exploit an RL agent to explore the design space automatically. The exploration aims to find the best trade-off, or one of the best possible trade-offs, between accuracy degradation and power/computation time.

\begin{figure}[tb]
\centering
\includegraphics[width=0.9\columnwidth]{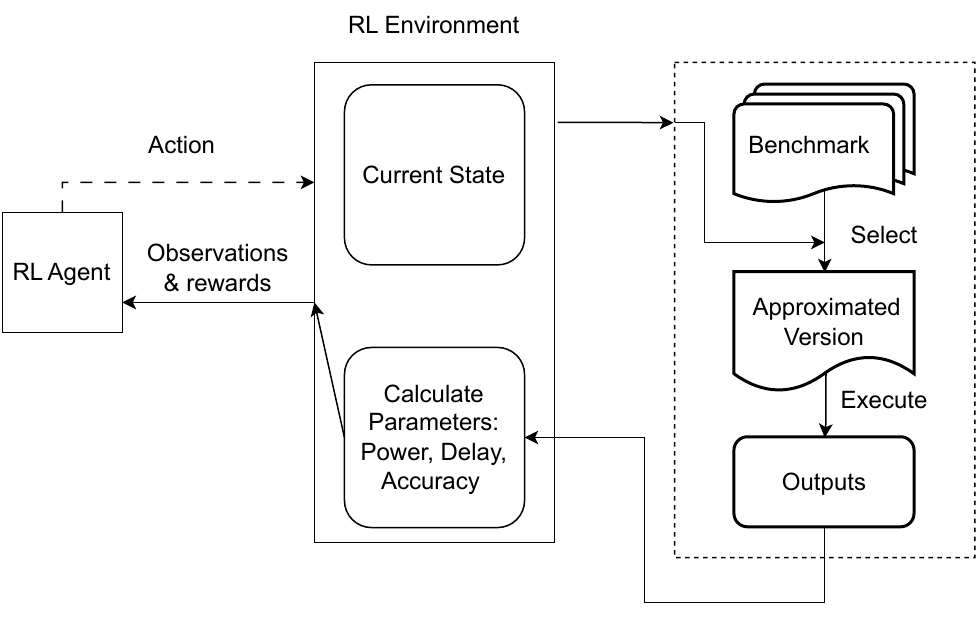}
  \caption{RL environment: at each step selects an approximated version of the benchmark to execute and calculates the parameters to return the new observations and rewards to the RL agent.}
  \label{fig:Gym_BenchExe}
\end{figure}

This process is depicted in \autoref{fig:Gym_BenchExe}. The environment uses available approximate operators and a set of variables to create an approximate application according to agent instructions. The approximate version's accuracy, power, and computation time are evaluated based on pre-characterized approximate operators. The agent receives observations and rewards from the environment, enabling them to take the next action to change the environment state. The RL setup mentioned above comprises four components: the environment, the agent, the state, and the action.

The \emph{environment} formally defined in \autoref{eq:env} is the system in which the agent operates. It models all the aspect of the optimization. First, the list of variables is modeled as a boolean vector where variables are indexed and selected if the corresponding index is 1 ($variables_{approx} = \{a_{0}, a_{1}, ..., a_{N-1} | a_{i} \in {0,1}\}$). The approximate adders ($adder \in \{1, 2, ..., N_{add}\}$) and the approximate multipliers ($multiplier \in \{1, 2, ..., N_{mul}\}$) are indexed to be linked to the set of available ones. Both sets are sorted by increasing accuracy degradation. Also the outcome of the computation is modeled as the decrease in accuracy, through an evaluation function comparing the output of the precise version with the approximate one ($\Delta acc = f_{err}(output_{precise}, output_{approx})$). In the actual implementation, $f_{err}$ evaluates the Mean Absolute Error (MAE) as the accuracy metric. MAE is calculated in \autoref{eq:MAE}, where $N$ is the number of the benchmark outputs. $exactOutput_i$ and $approxOutput_i$ are the exact and approximated values of each $output_i$. Eventually, the power reduction, expressed as the difference between the power consumption estimated for the precise execution and the approximated one ($\Delta power= power_{precise} - power_{approx}$) and the computational time reduction, expressed as the difference between the time used by precise execution in sums and multiplications and the one from the approximated version ($\Delta time = time_{precise} - time_{approx}$) are included.

\begin{multline}\label{eq:env}
environment = \{
adder, 
multiplier,
variables_{approx}, \\
{\Delta} acc,
{\Delta} power,
{\Delta} time\} 
\end{multline}


The \emph{state} represents the environment at a particular time (i.e., the list of approximated variables and the selected approximate adder and multiplier). This information is used to deploy the execution of an approximate application obtained through automatic code instrumentation that enables collecting observations (i.e., $\Delta acc$, $\Delta power$, and $\Delta time$). For example, the current state at a random step might look like $(4,5,(1,0,\cdots,0), 30, 100, 200)$, which, according to \autoref{eq:env}, states that the configuration includes approximate adder number 4 and multiplier number 5, and only the first variable in the list is selected for approximation. The last three elements represent observations meaning that, i.e., accuracy was reduced by 30 units, power consumption by 100 units (e.g., mW), and computation time was reduced by 200 units (e.g., ns). 

\begin{equation}\label{eq:MAE}
    MAE = \frac{1}{N}(\sum_{i=0}^{N-1} exactOutput_i - approxOutput_i)
\end{equation}


The \emph{agent} is the learning algorithm that interacts with the environment. It observes the current state of the environment and selects an action to take based on that state. The agent's goal is to learn a policy, which is a mapping from states to actions, that maximizes the expected cumulative reward.

The \emph{action} is the decision made by the agent at a particular time step to apply to the environment and change its state. In our work, three possible actions are: changing the type of the adder, changing the type of the multiplier, or removing/adding one variable in the list of approximated variables.\color{black}

In RL, the \emph{reward} helps the agent decide which action to take and learn the optimization policy. Our approach defines the reward as shown in \autoref{alg:RL_reward}. The reward obtained in each step is accumulated with previous rewards. If the total reward reaches a maximum predefined one, the agent stops.


\begin{algorithm}[tb]
\caption{RL Rewards at step \emph{i} }\label{alg:RL_reward}
{\scriptsize
Input: R  //Maximum cumulative reward \;
Input: R$_{cum}$ // Cumulative reward \;
Input: state := (adder, multiplier, variables, $\Delta$acc, $\Delta$power, $\Delta$time)\;
\If{$\Delta$acc $<=$ acc$_{th}$}{
    \If{(state$_{adder}$ == $N_{add}$ and state$_{mul}$ == $N_{mul}$ and variables contains all ones}{
        reward = R \;
        terminate = True\;
    }
    \Else{
        \If{($\Delta$power $>=$ $p_{th}$) and ($\Delta$time $>=$ $t_{th}$)}{
        reward = 1\;
        }
        \Else{
        reward = -1\;
        }
    }
}   
\Else{reward = -R\;}
R$_{cum}$ += reward\;
}
\end{algorithm}

Algorithm \ref{alg:RL_reward} analyzes the current state. A further approximation can be introduced if the accuracy loss is below the tolerable accuracy loss threshold for the benchmark (line 4). This acceptable threshold is an exploration parameter and can be adapted to the case. If the most approximated adder and multiplier are in use and all variables are approximated maximum reward possible is given, and the "terminate" flag is set to true (lines 5 to 8). Otherwise (lines 9 to 16), the gain in power consumption and computation time must be above a threshold to obtain a positive reward (+1). Otherwise, the agent gets a negative reward (-1). Line 19 gives the maximum negative reward if the accuracy loss is above its threshold. Finally, the cumulative reward is updated (line 21).

We used the Q-learning as a learning algorithm along with the RL environment. It is a model-free value-based RL algorithm that lets the agent learn the value of an action in each state~\cite{melo2001convergence}. Q-Learning finds the optimal action selection policy using a Q function that returns the expected future reward of an action at a state. In Q-learning, an accumulated reward can be defined, and the action selection is based on the state of the previous step beside the reward. Hence, it is suitable for maximizing the accumulated reward while considering the last state. For further details on how Q-function is updated, the reader may refer to~\cite{melo2001convergence}.

\section{Experimental Results}
\label{sec:results}

The RL implementation relies on the \texttt{Gymnasium}~\cite{Swaroop:2021aa} library, a fork of the OpenAI Gym~\cite{Brockman:2016aa} library, to implement the RL engine. \texttt{Gymnasium} is an open-source toolkit for developing and comparing RL algorithms. It provides an API to define environments, or simulated tasks, where agents can learn and improve their decision-making abilities through trial and error.
We implemented a Q-learning algorithm in which the maximum number of steps is 10,000. This number is selected upon trial and error. 
As an approximate component database, we exploited the EvoApproxLib~\cite{Mrazek:2017aa,Ceska:2017aa} that provides C models of 8 and 16 bits adders, and 8 and 32 bits multipliers. Tables \ref{tab:axc_add_op} and \ref{tab:axc_mul_op} report all selected operators and their Mean Relative Error Distance (MRED), power consumption, and computation time, ordered by MRED. 

\begin{table}[tb]
\caption{Selected adders from EvoApproxLib}
	\centering
	\begin{tabular}{|c|c|c|c|c|}
    \hline
       operator & Type & MRED & Power & Computation time \\ 
         &  &  &  (mW) & (ns) \\ \hline
        8-bit adder & 1HG & 0 & 0.033 & 0.63 \\ \hline
        8-bit adder & 6PT & 0.14 & 0.029 & 0.55 \\ \hline
        8-bit adder & 6R6 & 2.93 & 0.012 & 0.27 \\ \hline
        8-bit adder & 0TP & 6.16 & 0.0095 & 0.24 \\ \hline
        8-bit adder & 00M & 14.58 & 0.0046 & 0.17 \\ \hline
        8-bit adder & 02Y & 24.87 & 0.0015 & 0.11 \\ \hline
        16-bit adder & 1A5 & 0 & 0.072 & 1.28 \\ \hline
        16-bit adder & 0GN & 0.005 & 0.057 & 1.04 \\ \hline
        16-bit adder & 0BC & 0.018 & 0.051 & 0.95 \\ \hline
        16-bit adder & 0HE & 0.16 & 0.036 & 0.68 \\ \hline
        16-bit adder & 0SL & 9.54 & 0.011 & 0.27 \\ \hline
        16-bit adder & 067 & 22.35 & 0.0041 & 0.20 \\ \hline
    \end{tabular}
    \label{tab:axc_add_op}
\end{table}

\begin{table}[tb]
 \caption{Selected multipliers from EvoApproxLib}
	\centering
	\begin{tabular}{|c|c|c|c|c|}
    \hline
        operator & Type & MRED & Power & Computation time \\ 
         &  &  &  (mW) & (ns) \\\hline
        8-bit multiplier & 1JJQ & 0 & 0.391 & 1.43 \\ \hline
        8-bit multiplier & 4X5 & 0.033 & 0.380 & 1.40 \\ \hline
        8-bit multiplier & GTR & 1.23 & 0.303 & 1.46 \\ \hline
        8-bit multiplier & L93 & 4.52 & 0.178 & 1.11 \\ \hline
        8-bit multiplier & 18UH & 17.98 & 0.062 & 0.90 \\ \hline
        8-bit multiplier & 17MJ & 53.17 & 0.0041 & 0.11 \\ \hline
        32-bit multiplier & precise & 0 & 10.76 & 4.565 \\ \hline
        32-bit multiplier & 000 & 0.00 & 10.46 & 4.470 \\ \hline
        32-bit multiplier & 018 & 0.01 & 4.32 & 3.220 \\ \hline
        32-bit multiplier & 043 & 1.45 & 1.63 & 2.440 \\ \hline
        32-bit multiplier & 053 & 10.59 & 1.05 & 2.030 \\ \hline
        32-bit multiplier & 067 & 41.25 & 0.51 & 1.750 \\ \hline
    \end{tabular}
     \label{tab:axc_mul_op}
\end{table}

We tested our approach with two applications: Matrix Multiplication with two different matrices sizes (10 by 10 and 50 by 50) and FIR (with 100 and 200 samples, all white noise signals with Low Pass Filter functionality). The thresholds were set after executing the precise versions: the power ($p_{th}$) and computation time ($t_{th}$) thresholds were set to 50\% of their value for the precise version. Also, the precise outputs were averaged, and the accuracy threshold ($acc_{th}$) was set as 0.4 times the average output. 
Based on the rules defined in \autoref{alg:RL_reward}, the agent stopped after 2,000 and 4,000 steps for the Matrix Multiplication benchmarks and after 500 and 1,240 for the FIR benchmarks. All the results are reported in \autoref{tab:all_res}.

\begin{table}[htb]
\caption{ Explorations results for power, computation time, and accuracy}
	\centering
	\addtolength{\tabcolsep}{-3.5pt}
        \begin{tabular}{|l|c|c|c|c|}
    \hline
    \textbf{}&\multicolumn{2}{|c|}{\textbf{Matrix Mult.}}
\textbf{}&\multicolumn{2}{|c|}{\textbf{FIR}} \\
\cline{2-5} 
\textbf{Benchmarks} & \textit{10x10}& \textit{50x50}& \textit{100}&\textit{200} \\
\cline{4-5}
\hline \hline
		\multicolumn{5}{|c|}{$\Delta$ Power Consumption (mW)}\\
		\cline{1-5}
        min  & 15 & 0.55 & 529.515 & 1059.345 \\ \hline
        \textbf{solution} & 415.3 & 753.72 & 10850.855 & 1237.247 \\ \hline
        max & 418.4 & 1552.017 & 17344.390 & 34699.1 \\ \hline \hline
        \multicolumn{5}{|c|}{$\Delta$ Computation time (ns)}\\
		\cline{1-5}
        min& 50 & -90 & 563.135 & 1126.605 \\ \hline
        \textbf{solution} & 1780 & 1460.8 & 2664.385 & 3951.525 \\ \hline
        max & 1840 & 5707.6 & 6547.495 & 13098.89 \\ \hline \hline
        \multicolumn{5}{|c|}{Accuracy degradation}\\
		\cline{1-5}
        min  & 0.02 & 0 & 1096.03 & 395.74 \\ \hline
        \textbf{solution} & 19.95 & 0.736 & 1096.03 & 27580.345 \\ \hline
        max & 204.71 & 26.7964 & 31671.43 & 27580.35 \\ \hline \hline
        \multicolumn{5}{|c|}{Configuration}\\
		\cline{1-5}
        \textbf{Adder Type} & 00M & 6R6 & 0GN & 067 \\ \hline
        \textbf{Multiplier Type} & 17MJ & L93 & 043 & 018 \\ \hline
    \end{tabular}
    \label{tab:all_res}  
\end{table}

In Table \ref{tab:all_res}, power consumption and computation time are expressed as the difference ($\Delta$) between the parameter obtained from the precise version and the one obtained from the approximation run of the last step. The min and max rows correspond to the minimum, and maximum $\Delta$ found in the exploration. The accuracy degradation shows how much computation accuracy is reduced for the selected approximate version, using \autoref{eq:MAE}. The Adder and Multiplier types show which operators were selected for this approximate version, as named in Tables \ref{tab:axc_add_op} and \ref{tab:axc_mul_op}. 

For all cases, the agent identified approximations respecting the given constraints confirming the capability of exploring the space. Comparing the selected solution and the maximum observed values for the different parameters can indicate how the agent could push the approximation without violating the imposed constraints. 

\begin{figure}[hbt]
\centering
\includegraphics[width=0.9\columnwidth]{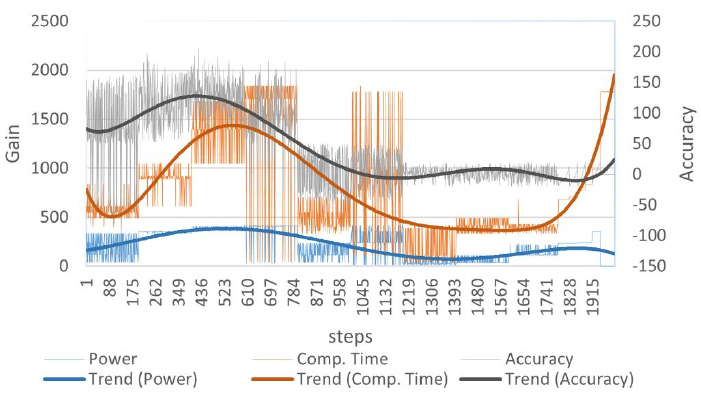}
  \caption{Exploration outcomes evolution through for Matrix Multiplication (10x10).}
  \label{fig:MM_plot}
\end{figure}

\begin{figure}[hbt]
\centering
\includegraphics[width=0.9\columnwidth]{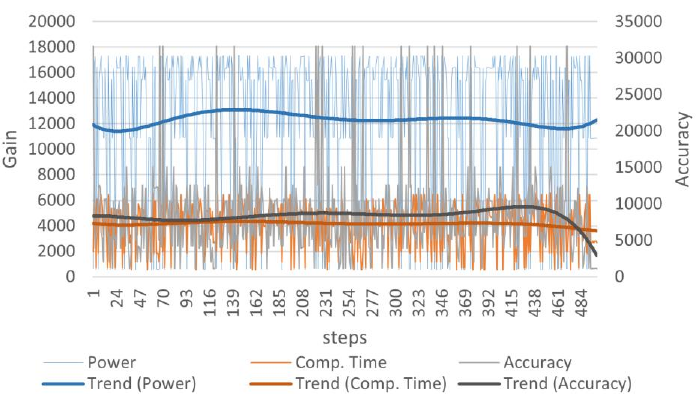}
  \caption{Exploration outcomes evolution through for FIR (100 samples).}
  \label{fig:FIR_plot}
\end{figure}

To better analyze the results, \autoref{fig:MM_plot} and \autoref{fig:FIR_plot} show the power and computation time alongside accuracy for all the exploration steps for the Matrix multiplication, 10 by 10, and FIR with 100 samples, respectively. The "Power" at each step represents the difference between the approximate and precise versions' power consumption. The same applies to the "Comp. Time" that shows the computation time. The trend lines help better visualize the exploration results. The different trends in the two figures highlight how the algorithm learns properly and move toward an optimization in the Matrix Multiplication benchmark, while it struggles in the FIR one.



\begin{figure}[hbt]
\centering
\includegraphics[width=0.9\columnwidth]{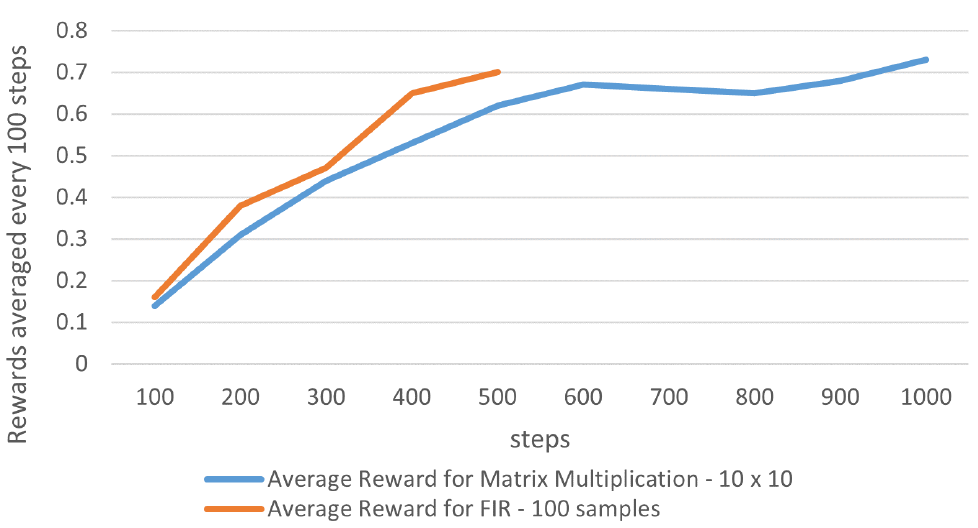}
  \caption{Average reward evolution for the Matrix multiplication (10x10) and FIR (100 samples).}
  \label{fig:reward_plot}
\end{figure}

\autoref{fig:reward_plot} shows the average reward every 100 steps for the same Matrix multiplication and FIR benchmark configurations. For Matrix multiplication, on average, the reward continuously improves in the first 600 steps, slowing down for a while and improving again, confirming that the agent learned. On the contrary, the FIR does not follow such a continuous improvement, ensuring that the learning strategy is not entirely effective and that further investigations are required.

\section{Conclusion}
\label{sec:conclusion}

In this paper, we proposed preliminary results obtained while developing a DSE strategy based on Reinforcement Learning to optimize an application, balancing the final accuracy with the power consumption and the computation time. 
Experimental results show that the agent can find solutions for a good trade-off between accuracy degradation and power and computation time reductions on some benchmarks. However, additional work is required to improve the learning strategy and make it general enough to work with a larger set of applications.  


\bibliographystyle{IEEEtran}
\bibliography{bibliography/IEEEabrv.bib, bibliography/biblio.bib}

\begin{thebibliography}{10}
\providecommand{\url}[1]{#1}
\csname url@samestyle\endcsname
\providecommand{\newblock}{\relax}
\providecommand{\bibinfo}[2]{#2}
\providecommand{\BIBentrySTDinterwordspacing}{\spaceskip=0pt\relax}
\providecommand{\BIBentryALTinterwordstretchfactor}{4}
\providecommand{\BIBentryALTinterwordspacing}{\spaceskip=\fontdimen2\font plus
\BIBentryALTinterwordstretchfactor\fontdimen3\font minus \fontdimen4\font\relax}
\providecommand{\BIBforeignlanguage}[2]{{%
\expandafter\ifx\csname l@#1\endcsname\relax
\typeout{** WARNING: IEEEtran.bst: No hyphenation pattern has been}%
\typeout{** loaded for the language `#1'. Using the pattern for}%
\typeout{** the default language instead.}%
\else
\language=\csname l@#1\endcsname
\fi
#2}}
\providecommand{\BIBdecl}{\relax}
\BIBdecl

\bibitem{Bosio:2022aa}
\BIBentryALTinterwordspacing
A.~Bosio, D.~M{\'{e}}nard, and O.~Sentieys, Eds., \emph{Approximate Computing Techniques}.\hskip 1em plus 0.5em minus 0.4em\relax Springer International Publishing, 2022. [Online]. Available: \url{https://doi.org/10.1007/978-3-030-94705-7}
\BIBentrySTDinterwordspacing

\bibitem{Hu:2019aa}
W.~Hu \emph{et~al.}, ``Exploring the design space of approximate arithmetic circuits using reinforcement learning,'' in \emph{2019 Design, Automation \& Test in Europe Conference \& Exhibition (DATE)}.\hskip 1em plus 0.5em minus 0.4em\relax IEEE, 2019, pp. 442--447.

\bibitem{Rizakis:2018aa}
M.~Rizakis \emph{et~al.}, ``Approximate fpga-based lstms under computation time constraints,'' in \emph{Applied Reconfigurable Computing. Architectures, Tools, and Applications}, N.~Voros \emph{et~al.}, Eds.\hskip 1em plus 0.5em minus 0.4em\relax Cham: Springer International Publishing, 2018, pp. 3--15.

\bibitem{Wu:2021aa}
Y.~Wu \emph{et~al.}, ``Ironman: Reinforcement learning based design space exploration for approximate computing,'' in \emph{2021 IEEE/ACM International Conference on Computer-Aided Design (ICCAD)}.\hskip 1em plus 0.5em minus 0.4em\relax IEEE, 2021, pp. 1--8.

\bibitem{gautier:2022sherlock}
Q.~Gautier \emph{et~al.}, ``Sherlock: A multi-objective design space exploration framework,'' \emph{ACM Transactions on Design Automation of Electronic Systems (TODAES)}, vol.~27, no.~4, pp. 1--20, 2022.

\bibitem{kaelbling1996rl}
L.~P. Kaelbling, M.~L. Littman, and A.~W. Moore, ``Reinforcement learning: A survey,'' \emph{Journal of artificial intelligence research}, vol.~4, pp. 237--285, 1996.

\bibitem{Savino:2019aa}
A.~Savino \emph{et~al.}, ``Approximate computing design exploration through data lifetime metrics,'' in \emph{2019 IEEE European Test Symposium (ETS)}, 2019, pp. 1--7.

\bibitem{Mrazek:2017aa}
V.~Mrazek \emph{et~al.}, ``Evoapprox8b: Library of approximate adders and multipliers for circuit design and benchmarking of approximation methods,'' in \emph{Design, Automation \& Test in Europe Conference \& Exhibition (DATE), 2017}, 2017, pp. 258--261.

\bibitem{Ceska:2017aa}
M.~{\v C}e{\v s}ka \emph{et~al.}, ``Approximating complex arithmetic circuits with formal error guarantees: 32-bit multipliers accomplished,'' in \emph{2017 IEEE/ACM International Conference on Computer-Aided Design (ICCAD)}, 2017, pp. 416--423.

\bibitem{melo2001convergence}
F.~S. Melo, ``Convergence of q-learning: A simple proof,'' \emph{Institute Of Systems and Robotics, Tech. Rep}, pp. 1--4, 2001.

\bibitem{Swaroop:2021aa}
S.~V. Swaroop \emph{et~al.}, ``Gymnasium,'' \url{https://github.com/s-vineet/gymnasium}, 2021, accessed: 2023-04-07.

\bibitem{Brockman:2016aa}
G.~Brockman \emph{et~al.}, ``Openai gym,'' \emph{arXiv preprint arXiv:1606.01540}, 2016.

\end{thebibliography}

\end{document}